\documentclass[letterpaper, 10 pt, conference]{ieeeconf}  

\IEEEoverridecommandlockouts                              

\usepackage{graphics} 
\usepackage{epsfig} 
\usepackage{lipsum}
\usepackage [english]{babel}
\usepackage [autostyle, english = american]{csquotes}
\MakeOuterQuote{"}
\usepackage{marvosym}
\usepackage{url}
\usepackage[nobiblatex]{xurl}

\title{\LARGE \bf
Annotating Covert Hazardous Driving Scenarios Online: Utilizing Drivers' Electroencephalography (EEG) Signals
}

\author{Chen Zheng, Muxiao Zi, Wenjie Jiang, Mengdi Chu, Yan Zhang, Jirui Yuan, Guyue Zhou and Jiangtao Gong\textsuperscript{\Letter}
\thanks{\textit{Corresponding author: Jiangtao Gong.}}
\thanks{The authors are with the Institute for AI Industry Research, Tsinghua University, Beijing, China. Emails:
        {\tt\small \{lastnamefirstname\}@air.tsinghua.edu.cn}}%
}

\begin{document}

\maketitle
\thispagestyle{empty}
\pagestyle{empty}

\begin{abstract}

As autonomous driving systems prevail, it is becoming increasingly critical that the systems learn from databases containing fine-grained driving scenarios. Most databases currently available are human-annotated; they are expensive, time-consuming, and subject to behavioral biases. In this paper, we provide initial evidence supporting a novel technique utilizing drivers' electroencephalography (EEG) signals to implicitly label hazardous driving scenarios while passively viewing recordings of real-road driving, thus sparing the need for manual annotation and avoiding human annotators' behavioral biases during explicit report. We conducted an 
EEG experiment using real-life and animated recordings of driving scenarios and asked participants to report danger explicitly whenever necessary. Behavioral results showed the participants tended to report danger only when overt hazards (e.g., a vehicle or a pedestrian appearing unexpectedly from behind an occlusion) were in view. By contrast, their EEG signals were enhanced at the sight of both an overt hazard and a covert hazard (e.g., an occlusion signalling possible appearance of a vehicle or a pedestrian from behind). Thus, EEG signals were more sensitive to driving hazards than explicit reports. Further, the Time-Series AI (TSAI,~\cite{tsai}) successfully classified EEG signals corresponding to overt and covert hazards. We discuss future steps necessary to materialize the technique in real life.

\end{abstract}

\section{INTRODUCTION}

Over the past few decades, self-driving cars have emerged as a promising mode of transportation, while security concerns are on the rise~\cite{400crashes, thedangers}. Central to resolving these concerns is hazard perception~\cite{sagberg2006hazard}, the vehicles' capacity to anticipate traffic hazards in driving situations, a concept borrowed from human drivers' driving competence.

There has been a growing body of literature proposing novel deep-learning models that allow cars to detect traffic hazards during autonomous driving~\cite{gupta2021deep, breitenstein2020systematization}. For example,~\cite{ramos2017detecting} presented a vision system fusing deep learning and geometric modelling to detect unexpected obstacles on the road. In~\cite{li2017traffic}, the authors proposed a new fully convolutional deep neural network architecture for semantic segmentation of traffic scenes on the pixel level, which adopted RGB-D photos as the input.~\cite{hadsell2008deep} introduced a learning-based approach for long-range vision that was capable of classifying complex terrain at distances up to the horizon, thus allowing high-level strategic planning.

Despite this hopeful trend, supervised deep-learning models in autonomous driving are data-thirsty, which, if not addressed properly, would reduce the models' precision and accuracy significantly. At the present stage, most, if not all, input images and videos are human-annotated, and human annotation is expensive, time-consuming, and contingent upon the validity and trustworthiness of human evaluation.

The International Organization for Standardization~\cite{iso_2022} has provided a framework on which human evaluation of safe driving can be based, titled "Road Vehicles - Safety of the Intended Functionality". Under this framework, driving scenarios were divided into four categories: known safe (Area 1), known unsafe (Area 2), unknown unsafe (Area 3), and unknown safe (Area 4). Increasing Area 1 and reducing Areas 2 and 3 would supposedly increase driving safety for an autonomous vehicle. For human annotators, however, it could be unnatural to explicitly label a driving scenario as unsafe if the unsafety is unknown (or absent from view). For example, a blind spot is unsafe because unexpected objects could appear from behind, but a driving scene featuring a blind spot only might be labelled as safe. A scenario involving a crash might be labelled as unsafe, but a near-miss situation might be labelled as safe.

One possibility to get around this impasse would be to utilize the annotators' physiological signals and mark up the unsafe scenarios implicitly. Previous studies have been conducted analyzing the power spectral density features of drivers' EEG signals upon exposure to a hazard cue (occlusion) during simulated driving, e.g.,~\cite{guo2020recognizing}. Power spectral density analysis has also been widely used in studying fatigue driving, distracted driving, and emotional driving~\cite{peng2022application}. However, power spectral density analysis does not provide the precise time by which the drivers have detected a hazard. On the contrary, event-related potentials (ERPs), known as deflections in the EEG waveform with a positive or negative polarity upon detection of the stimuli, are highly temporally sensitive to the onset of traffic hazards~\cite{li_chang_sui_2022},~\cite{li2022exploratory} and warning signs~\cite{ma_bai_pei_xu_2018, zhu_ma_bai_hu_2020}.

In this regard, we propose an annotation system that differentiates between safe and unsafe driving scenarios based on driver's EEG signals, in particular, the P400 and N500 ERP components. The P400 component is a positive voltage deflection that peaks around 400 ms post stimulus onset, and the N500 is a negative voltage deflection that peaks around 500 ms post stimulus onset. We focused on EEG signals recorded by electrodes placed above the premotor and motor cortex (lying in the frontal lobe), which are involved in action planning and organization (i.e., the FPz, AF4, F4, AF3, F3, and F1). The frontal P400 has been found triggered by attention engagement, e.g.,~\cite{guy2016cortical}, and the N500 component by unpredictability, e.g.,~\cite{polezzi2008predicting}. The most prominent N500 effect seems to occur over the frontal areas~\cite{de1999effect}.

\begin{figure*}
\centering
  \includegraphics[width=1\textwidth]{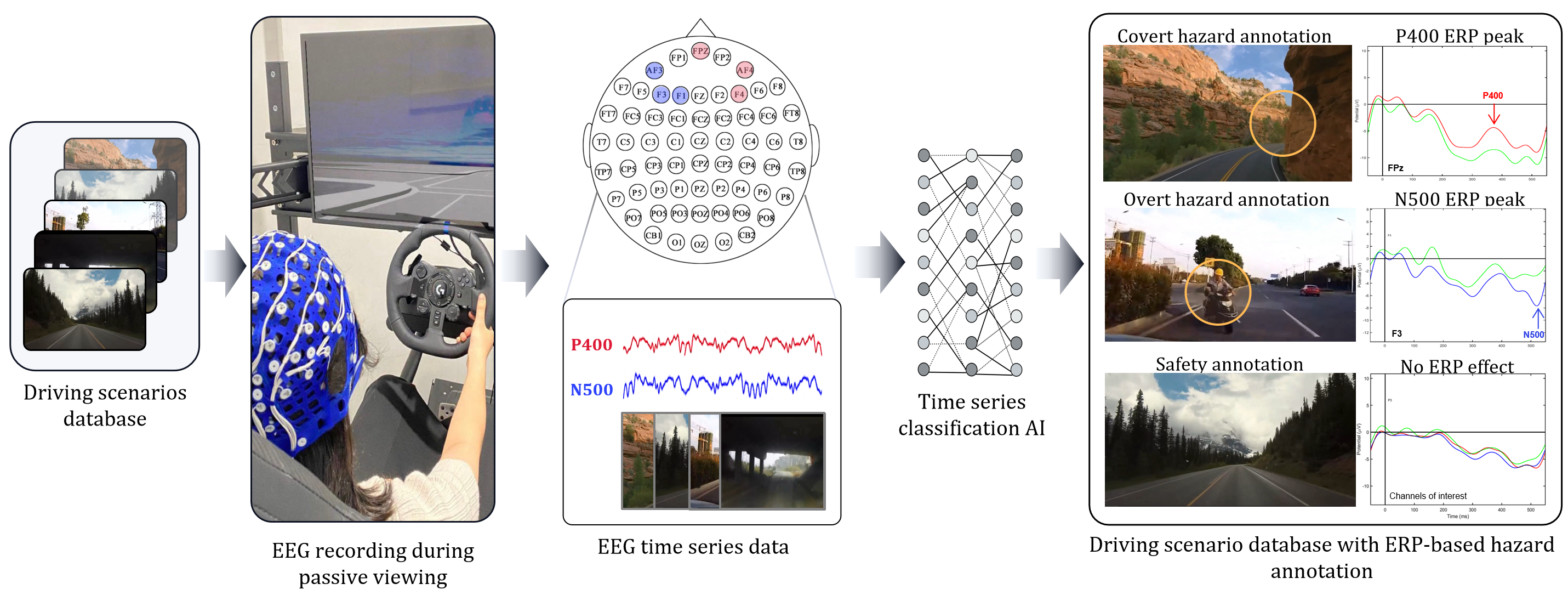}
  \caption{This study provides initial evidence supporting implicit annotation of overt and covert traffic hazards based on drivers' ERPs during passive viewing of recordings of driving scenarios.}
  \label{fig:ICRA2023_example}
\end{figure*}

Towards this end, we conducted two experiments showing participants real-life and animated recordings of driving scenarios. We invited experienced driving instructors as participants and instructed them to report danger whenever they detected it while viewing the recordings (Fig. 1.). We recorded the participants' EEG activities as they were performing the task. The results showed that drivers tended to report danger only when a hazardous vehicle or pedestrian was featured in the scene; they could either appear unexpectedly from behind an occlusion or were visible since the beginning of the video clip, and the occlusion itself would not suffice to trigger an explicit report. However, when we fit single-trial ERP data to a deep-learning model (Time Series AI, TSAI,~\cite{tsai}) designed for time-series classification, the model successfully differentiated between driving scenes that did and did not feature an occlusion, which means the occlusion made a difference to the participants' ERPs as compared to when the road was unobstructed. Additionally, the model successfully classified driving scenes that did and did not feature an overt hazard (i.e., a pedestrian) as well. As vehicles, pedestrians, and occlusions are all traffic hazards, these results suggested the participants' EEG signals were more sensitive to hazardous scenes than their explicit report, thus supporting the possibility of utilizing the drivers' EEG signals to annotate hazardous driving scenes implicitly at the absence of explicit report.

The contributions of this study are: a) We identify two ERP components, namely, the P400 and N500, which are characteristic of a driver's hazard detection in a driving scenario. b) By fitting single-trial ERP data to a deep-learning model, we provide evidence that hazardous and safe driving scenes could be differentiated through Time Series Classification (TSC) on the basis of the P400 and N500 ERP components they elicit. c) We propose a new technique for hazard annotation that utilizes drivers' physiological signals, particularly their ERPs; this technique would enable drivers to annotate hazardous driving scenes during passive viewing and mark up covert hazards implicitly, thus increasing the efficiency of human annotation and expanding the pool of known unsafe scenarios. 

\section{METHODOLOGY}

\subsection{Participants}

We invited 10 experienced driving instructors as participants according to the following criteria: a) having over 10 years of experience as a driving instructor, b) having a valid Chinese driver's license, c) having normal or corrected-to-normal eye vision, and d) having no prior knowledge about video annotation or EEG. All participants were right-handed males with an age range between 30 and 51 (\textit{M} = 42.91, \textit{SD} = 6.06). No participant reported a history of neurological or psychiatric illness. The study was approved by Tsinghua University's Internal Review Board. 

\begin{figure}
\centering
  \includegraphics[width=0.7\columnwidth]{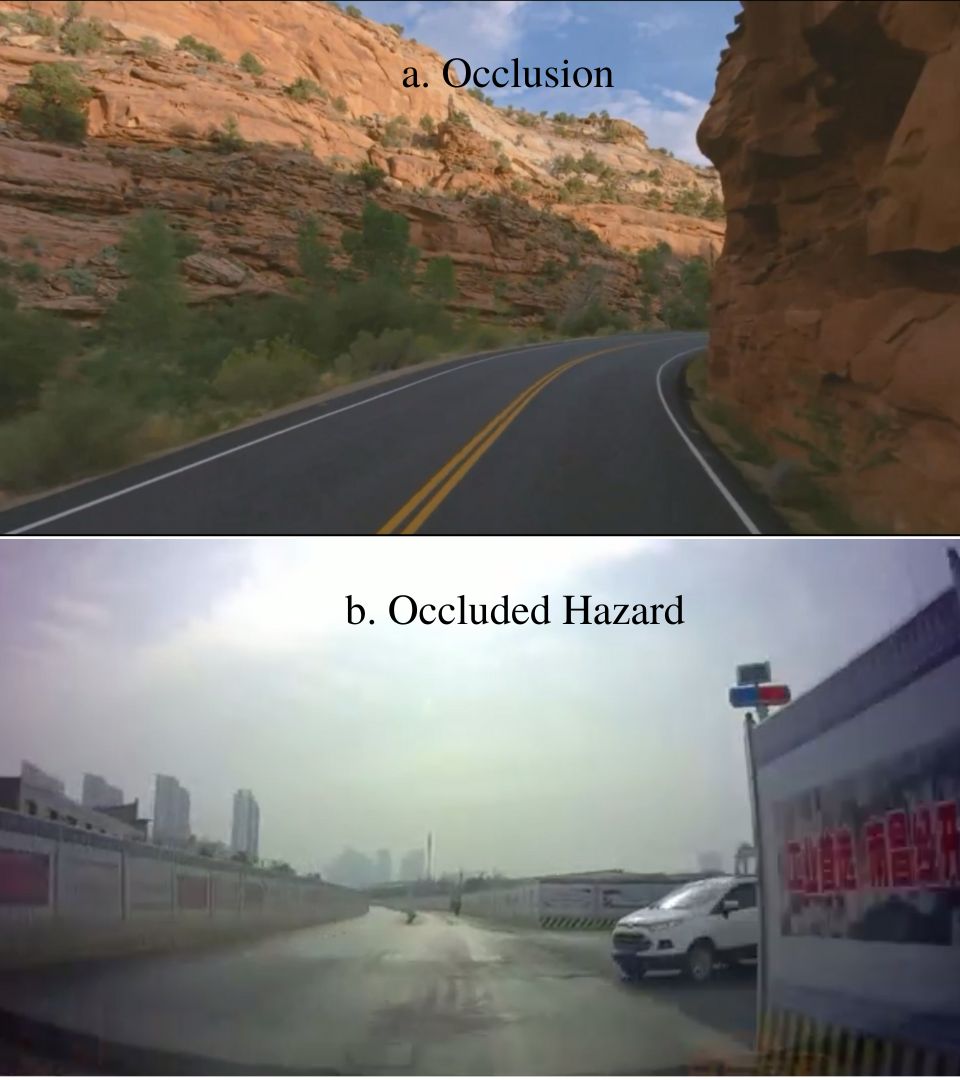}
  \caption{Two conditions in the first experiment.}
  \label{fig:ICRA2023_example}
\end{figure}

\begin{figure}
\centering
  \includegraphics[width=0.9\columnwidth]{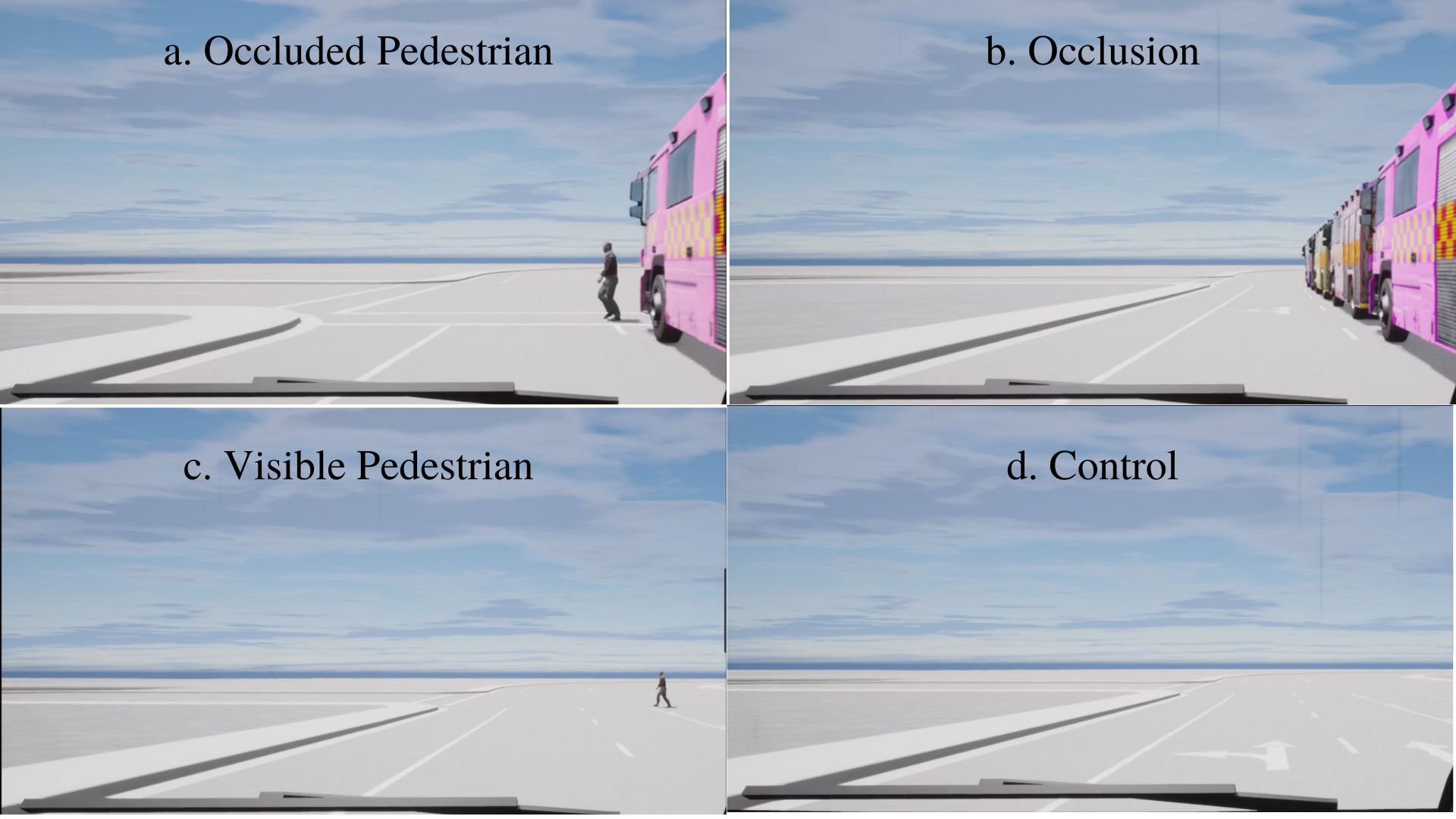}
  \caption{Four conditions in the second experiment.}
  \label{fig:ICRA2023_example}
\end{figure}

\subsection{Experimental Design and Stimuli}

The study comprised two experiments. The first experiment aimed at investigating the scenarios in which the participants would explicitly report danger, and the second served to identify the ERPs characteristic of these scenarios. 

The first experiment had a one-way within-subjects design. The variable was the type of hazard a video featured, that is, whether the videos featured an occlusion (the \textit{occlusion} condition) or a vehicle appearing from behind the occlusion (the \textit{occluded hazard} condition). We used recordings of real-life driving situations obtained from~\cite{gan2021constructing} and YouTube; see Appendix A for a link to the YouTube playlist. As in Fig. 2., videos in the \textit{occlusion} condition featured a straight and level pathway with an occlusion (e.g., an intersection, a turning point, etc.) at the end of it, and the ego car was the only automobile in view. Videos in the \textit{occluded hazard} condition featured an unexpected vehicle appearing from behind an occlusion, and the hazard vehicle was the only automobile aside from the ego car. Videos in the \textit{occlusion} condition were five seconds long, and those in the \textit{occluded hazard} condition lasted for a random length of time before the hazard appeared and two seconds after.

The second experiment also had a one-way within-subjects design. The variable was the same, but, this time, had four levels: \textit{occluded pedestrian}, \textit{occlusion}, \textit{visible pedestrian}, and \textit{control} (Fig. 3.). Videos in the \textit{occluded pedestrian} condition featured a pedestrian appearing from behind an array of buses (i.e., the occlusion) on a crossroad. Videos in the \textit{occlusion} condition featured the buses on the crossroad. Videos in the \textit{occluded pedestrian} condition featured the pedestrian crossing the crossroad. The \textit{control} condition featured the crossroad with an unobstructed view. The videos were created using CARLA~\cite{dosovitskiy2017carla}. The colors of the buses and the appearance of the pedestrian were randomized, and the left-right position of the occlusion as well as the walking direction of the pedestrian were counterbalanced across trials. The occlusion (if featured) was visible since video onset, and the pedestrian (if featured) appeared five seconds post video onset. All videos were seven seconds in length and had the same parametric setting.

In both experiments, the video clips were taken from the driver's first-person perspective, and the participants viewed the clips in random orders.

\subsection{Apparatus}

\subsubsection{Car Driving Simulator and Stimuli Presentation}

Both experiments took place in a Logitech G923 driving simulator, which consists of a driving seat and a steering wheel (Fig. 1.). The steering wheel had two shifter paddles positioned behind the left and right wheel spokes. The stimuli were presented on a flat, black 32-inch HPC monitor (710 * 420 * 80 mm) with 1920 * 1080 pixels and a refresh rate of 75Hz. The screen was placed 60 to 80 cm from the participants' eyes. During the experiments, the  stimuli were presented using the E-Prime 3.0 software~\cite{eprime} in a 16:9 aspect ratio with a resolution of 1080 pixels. For the sake of signal quality, we did not specify a fixed viewing distance or angle and ensured the participants were comfortably seated.

\subsubsection{EEG Acquisition}
We recorded the participants' EEG signals continuously (1000 Hz sampling rate) during each experiment using a 64-channel Ag/AgCl-electrode Neuroscan SynAmps² Model 8050 Quik-Cap EEG cap and the CURRY 8 X Data Acquisition package. The EEG electrodes were placed according to a modified International 10-10 system. The EEG recordings were amplified using the Neuvo 64-channel amplifier. The input impedance was 250 k$\Omega$, and we kept the electrode impedance below 10 k$\Omega$ during the experiments. EEG was initially acquired using the REF electrode as a reference and re-referenced to the M1 and M2 electrodes during preprocessing.

\subsection{Procedure and Experimental Task}

The participants underwent the experiments one at a time in a row; they were encouraged to take a break in between and start the second experiment at will.

On the day of the experiments, the participants signed the consent form and filled in some demographic information, then seated themselves in the driving simulator (Fig. 1.) after washing and drying their hair. 

Each condition in each experiment had 12 practice trials and 20 experimental trials. Each trial (Fig. 4 and Fig. 5.) began with a black cross presented at the center of the screen, and the participants were instructed to focus their attention at the intersection of the bars. Each cross lasted 800 to 1000 ms at random before the video clip started to play. The participants pressed the right shifter paddle to indicate detection of a hazard in a video clip. The video clip would stop playing as soon as the shifter paddle was pressed; otherwise, it would keep playing until its full length. Then the participants saw an empty screen that lasted 1000 ms, after which the subsequent trial would begin. 
Whenever a hazardous vehicle or a pedestrian was featured in a video clip, the ego vehicle would hit them if the participant did not press the shifter paddle.

\begin{figure}
\centering
  \includegraphics[width=0.9\columnwidth]{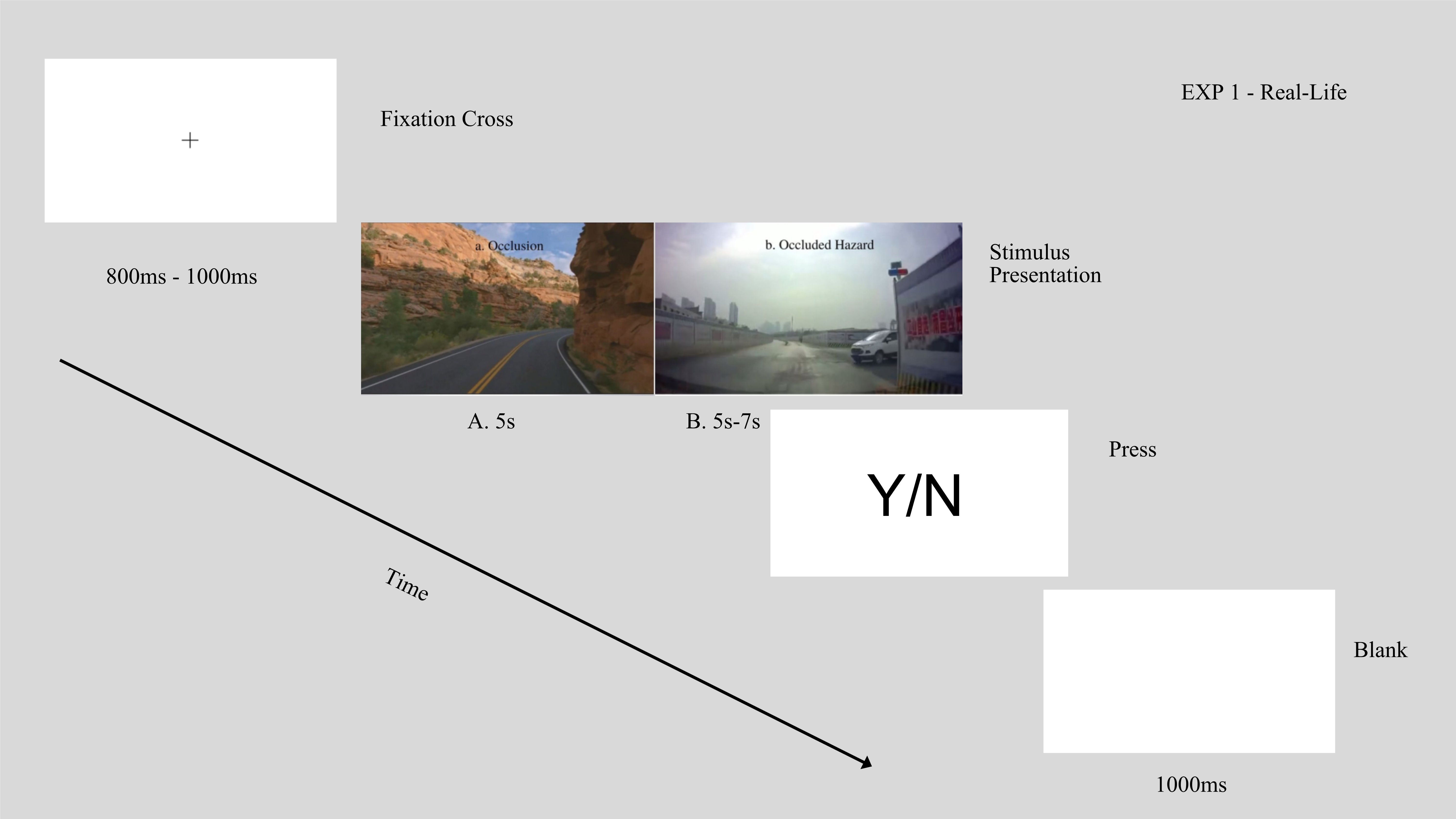}
  \caption{Procedure of the first experiment using real-life driving scenes.}
  \label{fig:ICRA2023_example}
\end{figure}

\begin{figure}
\centering
  \includegraphics[width=0.9\columnwidth]{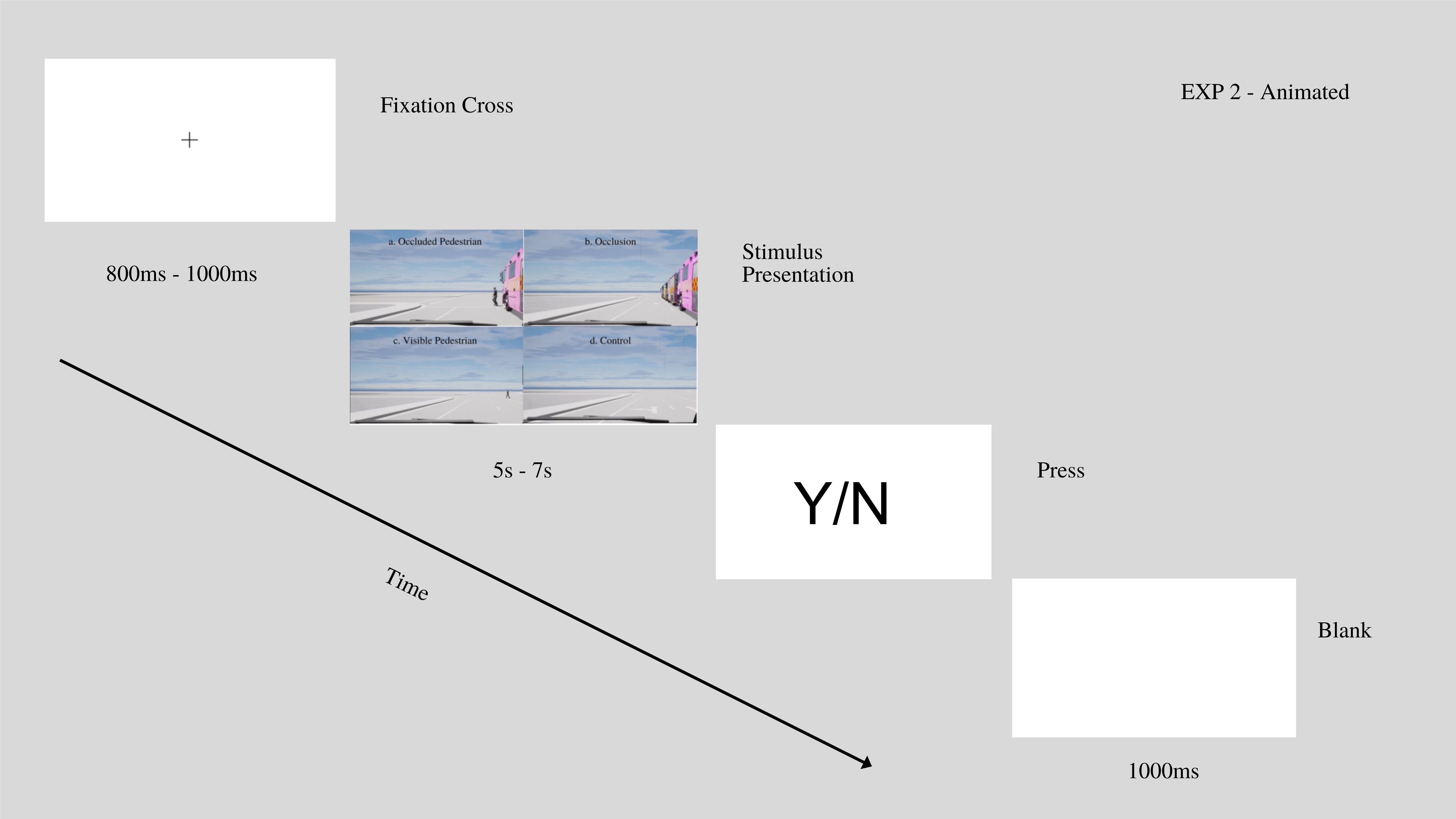}
  \caption{Procedure of the second experiment using animated driving scenes.}
  \label{fig:ICRA2023_example}
\end{figure}

\subsection{Data Analysis}

\subsubsection{Behavioral Data}
In each condition, we counted the number of trials in which the participants pressed the shifter paddle. Then, for each experiment, we used SPSS 27 to conduct Chi-square analysis to test the hypothesis that participants were more likely to report detection of danger when it was overt than when it was covert or absent. Overt danger was a vehicle or pedestrian who either appeared from behind the occlusion (i.e., the \textit{occluded hazard} condition in the first experiment and the \textit{occluded pedestrian} condition in the second experiment) or were visible since the beginning of a video clip (i.e., the \textit{visible pedestrian} condition in the second experiment). Covert danger was an occlusion from behind which a vehicle or pedestrian might or might not appear (i.e., the \textit{occlusion} conditions in the two experiments). The \textit{control} condition in the second experiment featured no danger in the scene.

\subsubsection{EEG Preprocessing}

We used EEGLAB~\cite{delorme2004eeglab} for EEG data preprocessing. We re-referenced the data to the M1 and M2 electrodes and band-filtered them at 0.1 to 40 Hz. We then interpolated the bad channels identified for each participant, respectively. We conducted independent component analysis (ICA) and removed signal noises resulting from eye movements, channel noise, heartbeat, and limb movement. Finally, we extracted epochs for each time-locking event type in the second experiment at -500 to 600 ms in relation to video onset. For each epoch extracted, we subtracted the 500-millisecond pre-stimulus interval (-500 to 0 ms) from each post-stimulus time point to correct for baseline differences. 

\subsubsection{Time Series Classification Algorithm}
For epochs extracted from conditions in the second experiment, we fit the data into Time Series AI (TSAI,~\cite{tsai}) to classify a) the P400 ERP component as from either the \textit{control} condition or the \textit{occluded pedestrian} and the \textit{occlusion} conditions combined and b) the N500 component as from either the \textit{control} or the \textit{occluded pedestrian} condition. We used 320 (16 per condition per participant) time series (i.e., wave amplitudes sampled at a rate of 1000 Hz from 351 to 450 ms for the P400 component and from 451 to 550 ms for the N500 component) as the training dataset. The trained model was then applied to the test dataset, which included the remaining 80 time series (four per condition per participant) to see how well it could correctly classify different types of EEG signals.

We trained the model separately for each electrode of interest. For the P400 component, we trained the electrodes FPz, AF4, and F4, and for the N500 component, we trained the electrodes AF3, F1, and F3.

\section{RESULTS}

\begin{figure*}
\centering
  \includegraphics[width=0.9\textwidth]{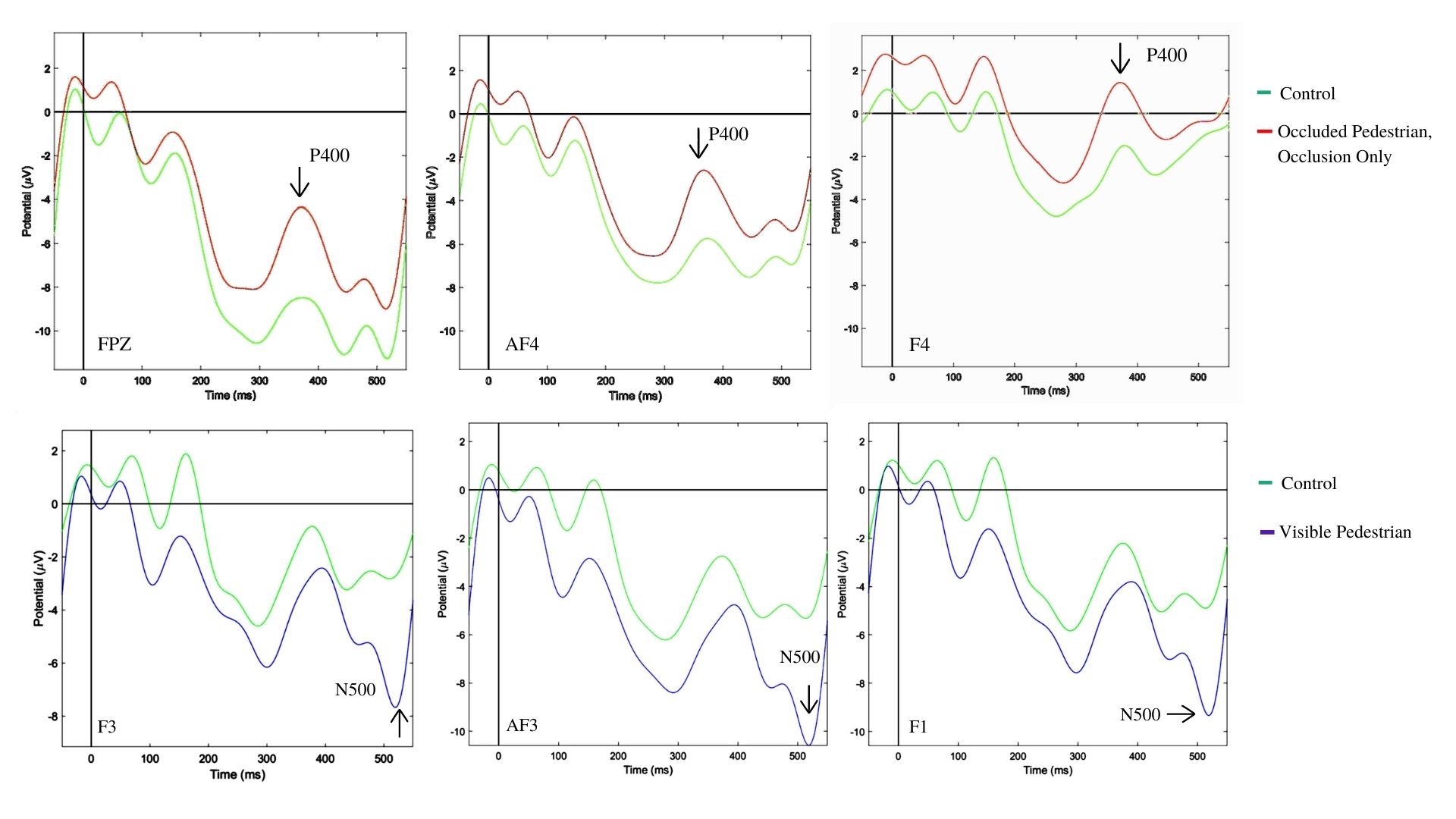}
  \caption{The P400 and N500 components upon detection of a driving hazard.}
  \label{fig:ICRA2023_example}
\end{figure*}

\subsection{Behavioral Results}
According to the results of a Chi-square test, in the first experiment, the way participants pressed the shifter paddle differed between the \textit{occlusion} and \textit{occluded hazard} conditions ($\chi^2$(1, \textit{N} = 440) = 361.69, \textit{p} $<$ .001). When a hazard was present, the participants pressed the shifter paddle 217 times out of 220 trials. By contrast, when a hazard was absent (and only the occlusion was present), the participants pressed the shifter paddle 18 times out of 220 trials.

A similar result was obtained in the second experiment ($\chi^2$(3, \textit{N} = 880) = 571.81, \textit{p} $<$ .001). When a video featured a pedestrian appearing from behind an occlusion and crossing the road, the participants pressed the shifter paddle 220 and 219 times out of 220 trials, respectively. When no pedestrian was present, that is, when only the occlusion was present and when no occlusion or pedestrian was present, the participants pressed the shifter paddle 59 times and 35 times out of 220 trials, respectively.

\subsection{Characteristic ERP Components}

Fig. 6. shows waveforms of the electrodes FPz, AF4, F4, F3, AF3, and F1 during the window of 50 ms before and 550 ms after video onset. We picked the time windows of 351-450 ms and 451-550 ms, respectively for the P400 and N500 components.

FPz showed an enlarged amplitude of the P400 component upon detection of occlusion as compared to the control condition (\textit{t}(9) = 3.51, \textit{p} = .007, \textit{d} = 1.11). This difference was non-significant for the electrodes AF4 (\textit{t}(9) = 1.59, \textit{p} = .147, \textit{d} = 0.50) and F4 (\textit{t}(9) = 1.06, \textit{p} = .316, \textit{d} = 0.34). AF3 showed an enlarged amplitude of the N500 component upon detection of a pedestrian as compared to the control condition (\textit{t}(9) = 2.64, \textit{p} = .027, \textit{d} = 0.84). This difference was borderline significant for the electrodes F3 (\textit{t}(9) = 2.28, \textit{p} = .049, \textit{d} = 0.72) and F1 (\textit{t}(9) = 2.31, \textit{p} = .046, \textit{d} = 0.73). These results suggested that, on a statistical level, the participants' brain activity was enhanced at FPz, AF3, F3, and F1 when they detected overt and covert danger in a driving scene.

\subsection{Algorithm for EEG Time Series Classification}

Based on the P400 component recorded at the FPz, AF4, and F4 electrodes, the TSAI classified the driving scenes in the second experiment as featuring an occlusion vs. control with an accuracy of .61, .63, and .61, respectively. Based on the N500 component recorded at the AF3, F3, and F1 electrodes, the TSAI classified the driving scenes in the second experiment as featuring a visible pedestrian vs. control with an accuracy of .74, .64, and .68, respectively.

\section{DISCUSSION}

The study set out with the aim of proposing an automated system that annotates hazardous driving scenes based on drivers' ERPs while they passively view recordings of driving scenarios. We identified two ERP components, namely, the P400 and N500, that EEG devices could capture 400 and 500 ms after the onset of a hazardous scene. At the electrode FPz, the participants' P400 amplitudes increased when they detected an occlusion in the driving scene as compared to when the scene was unobstructed, and at the electrodes AF3, F3, and F1, their N500 amplitudes increased when they saw a pedestrian crossing the road as compared to when the road was empty. We then fit single-trial EEG data to TSAI, a deep-learning model designed for time series classification. The model successfully differentiated between driving scenes that did and did not feature an occlusion based on the P400 component recorded at the electrodes FPz, AF4, and F4 respectively, as well as those that did and did not feature a pedestrian based on the N500 component recorded at the electrodes AF3, F3, and F1 respectively.

Note that we hereby define a \textit{hazardous scenario} as including both overt and covert hazards. In this study, overt hazards were vehicles and pedestrians either crossing the road or appearing unexpectedly from behind an occlusion, and covert hazards were buildings, trees, arrays of buses, etc., serving as the occlusion. Thus, increased N500 amplitudes would suggest enhanced brain activity upon detection of an overt hazard, and increased P400 amplitudes would suggest enhanced brain activity upon detection of a covert hazard. 

By contrast, on the behavioral level, the participants tended to only report their detection of overt hazards and not covert hazards. In this sense, the participants' EEG signals seemed more sensitive to hazardous driving scenarios than their explicit report. 

These results are of particular importance in near-miss scenarios where injuries and damages are avoided by a slight shift in space or time. The hazards in near-miss scenarios are covert, but the situations are as dangerous as actual accidents. To ensure driving safety, it would be critical for autonomous driving systems to detect near-miss scenarios and react accordingly. In terms of databases of near-miss scenarios for autonomous driving, ~\cite{kataoka2018drive} and~\cite{kataoka2020joint} provided two of the first databases consisting of near-miss incidents, but due to the inherent rarity of near-miss cases and the difficulty of capturing abnormal visual features (e.g., those in avoiding movements)~\cite{kataoka2020joint}, these datasets have yet to reach the scale of databases comprising normal real-world traffic such as KITTI~\cite{geiger2012kitti} and CitySpaces~\cite{cordts2016cityscapes} (which, on the other hand, do not contain near-miss scenarios).

Utilizing drivers' ERPs during passive viewing would help capture near-miss scenarios as well as other dangerous driving scenarios characterized by covert hazards that are previously unknown. It would also help reduce, if not spare, the workload of manual annotation of such recordings because ERP data is in the form of time series and could be captured automatically by time series classification algorithms. 

However, to materialize the technique in real-life driving situations, many remaining questions must be answered. First and foremost, would the P400 and N500 effects hold in other driving scenarios, for example, when the driving scene were complicated and composed of multiple vehicles and pedestrians? Second, would the effects hold if the drivers adopted a more natural pattern of response to hazards, for example, by steering, braking, or releasing the throttle in response to an emergency? Third, how to increase the accuracy of EEG time-series classification when the signals are noisy, especially when the recordings feature real-life driving scenarios and TSAI might no longer be suitable?

To answer the first two questions, according to the International 10-10 System of the placement of EEG electrodes, the electrodes analyzed in this study were placed over the participants' prefrontal cortices. Specifically, the FPz was placed over the anterior prefrontal cortex (also known as the frontopolar prefrontal cortex and the rostrolateral prefrontal cortex), the AF4 and the AF3 were placed over the dorsolateral prefrontal cortex, and the F4, the F3, and the F1 over the frontal eye fields. 

Neuroscience researchers have proffered many hypotheses for the functions of the anterior prefrontal cortex. For example, one of the major hypotheses~\cite{koechlin2007anterior} is that the anterior prefrontal area performs a domain-general function in coordinating multiple simultaneous cognitive processes implicated in complex behaviors. A similar argument in~\cite{gilbert2006functional} contends that the anterior prefrontal cortex is involved in working memory and multi-task coordination. Thus, activity in the anterior prefrontal cortex should increase when drivers comprehend and assess recordings of more complicated driving scenarios.

The dorsolateral prefrontal cortex is widely involved in various cognitive processes. Of special interest is its role in planning~\cite{fincham2002neural} and intention attribution~\cite{goel1995modeling, goel1997seats}. In the second experiment, When the participants saw a covert hazard (i.e., the occlusion), they must get ready to press the shifter paddle as soon as the overt hazard (i.e., a pedestrian) appeared. When the overt hazard, a pedestrian, was visible since the beginning, the participants must infer the pedestrian's intention to decide how likely the car might hit them. Thus, action planning and intention attribution seemed independent from the way that participants responded to the hazards, that is, planning and intention attribution would take place regardless of whether the participants pressed the shifter paddle, steered, braked, or released the throttle.

Finally, activity in the frontal eye fields correlates with the level of uncertainty one must cope with~\cite{volz2005variants}. In the second experiment of this study, covert traffic hazards were the source of uncertainty, as the participants did not know whether or not an overt hazard (a pedestrian) would appear at the end of the covert hazard (the occlusion), nor could they predict the consequence of an overt hazard as soon as it appeared on the screen. As real-life driving situations would involve more diverse types of uncertainty, we would expect the frontal eye fields to be more activated if drivers were exposed to more complicated driving scenes.

In this sense, it seems the anterior prefrontal cortex, the dorsolateral prefrontal cortex, and the frontal eye fields are implicated in the detection of and response to covert as well as overt hazards in general, which falls in line with previous findings employing functional Magnetic Resonance Imaging (fMRI)~\cite{gharib2020neural}. Thus, it is reasonable to argue that the P400 and N500 effects in the detected prefrontal areas should hold in complicated driving scenarios in real life and regardless of the driver's way of responding.

However, to acquire more solid evidence, more laboratory EEG studies shoud be conducted employing finer-grained real-life driving recordings as well as more realistic animations. The participants should also be allowed to respond as they would on the road. This would inevitably increase the noise captured by EEG, for example, signals resulting from sudden movements. In this case, one possibility to increase the accuracy of EEG time-series classification would be to collect multi-modal data during the experiment, for example, drivers' hand and foot movement, eye-fixation, electrodermal activities, etc., which would also provide a temporal reference against which researchers could carefully select the sections of EEG data to analyze. 

\section{CONCLUSIONS}

When it comes to annotating hazardous driving scenarios, annotators' physiological signals (e.g., EEG) are sensitive to covert hazards which are hard to extract manually. We thus propose a novel annotation technique utilizing annotators' EEG signals, specifically, the P400 and N500 ERP components captured in the prefrontal electrodes. In a controlled experiment, the annotators' (driving instructors) P400 and N500 amplitudes increased upon perception of covert and overt hazards as compared to safe conditions. The increase was successfully detected by a deep-learning model (i.e., the TSAI) designed for time-series classification. More data would be needed to increase the ecological validity of the ERP effects as well as the algorithm's classification accuracy. We argue that this technique would help avoid human behavioral biases during annotation and reduce the cost and workload of human annotation.

\section*{ACKNOWLEDGEMENT}

This work was supported by Baidu, Inc.

\addtolength{\textheight}{0cm}   



\section*{APPENDIX}

Appendix A: Playlist of driving scenarios in the \textit{Occlusion} and \textit{Occluded Hazard} conditions in the first experiment obtained from YouTube

\url{https://www.youtube.com/playlist?list=PLvmIQa5eOlV-GJTlL1JS3jNluSWKbTuuw}

\bibliographystyle{IEEEtran}
\bibliography{IEEEfull,root}

\begin{thebibliography}{10}
\providecommand{\url}[1]{#1}
\csname url@rmstyle\endcsname
\providecommand{\newblock}{\relax}
\providecommand{\bibinfo}[2]{#2}
\providecommand\BIBentrySTDinterwordspacing{\spaceskip=0pt\relax}
\providecommand\BIBentryALTinterwordstretchfactor{4}
\providecommand\BIBentryALTinterwordspacing{\spaceskip=\fontdimen2\font plus
\BIBentryALTinterwordstretchfactor\fontdimen3\font minus
  \fontdimen4\font\relax}
\providecommand\BIBforeignlanguage[2]{{%
\expandafter\ifx\csname l@#1\endcsname\relax
\typeout{** WARNING: IEEEtran.bst: No hyphenation pattern has been}%
\typeout{** loaded for the language `#1'. Using the pattern for}%
\typeout{** the default language instead.}%
\else
\language=\csname l@#1\endcsname
\fi
#2}}

\bibitem{tsai}
\BIBentryALTinterwordspacing
I.~Oguiza, ``tsai - a state-of-the-art deep learning library for time series
  and sequential data,'' Github, 2022. [Online]. Available:
  \url{https://github.com/timeseriesAI/tsai}
\BIBentrySTDinterwordspacing

\bibitem{400crashes}
{The Associated Press}, ``Nearly 400 car crashes in 11 months involved
  automated tech, companies tell regulators,'' National Public Radio.
  \url{https://www.npr.org/2022/06/15/1105252793/nearly-400-car-crashes-in-11-months-involved-automated-tech-companies-tell-regul}
  (accessed Sep. 1, 2022).

\bibitem{thedangers}
{Clifford Law Offices PC}, ``The dangers of driverless cars,'' {National Law
  Review.} \url{https://www.natlawreview.com/article/dangers-driverless-cars}{
  (accessed Sep. 1, 2022)}.

\bibitem{sagberg2006hazard}
F.~Sagberg and T.~Bj{\o}rnskau, ``Hazard perception and driving experience
  among novice drivers,'' \emph{Accident Analysis $\&$ Prevention}, vol.~38,
  no.~2, pp. 407--414, 2006.

\bibitem{gupta2021deep}
A.~Gupta, A.~Anpalagan, L.~Guan, and A.~S. Khwaja, ``Deep learning for object
  detection and scene perception in self-driving cars: Survey, challenges, and
  open issues,'' \emph{Array}, vol.~10, 2021, 100057.

\bibitem{breitenstein2020systematization}
J.~Breitenstein, J.-A. Term{\"o}hlen, D.~Lipinski, and T.~Fingscheidt,
  ``Systematization of corner cases for visual perception in automated
  driving,'' in \emph{IEEE Intelligent Vehicles Symposium (IV)}.\hskip 1em plus
  0.5em minus 0.4em\relax IEEE, 2020, pp. 1257--1264.

\bibitem{ramos2017detecting}
S.~Ramos, S.~Gehrig, P.~Pinggera, U.~Franke, and C.~Rother, ``Detecting
  unexpected obstacles for self-driving cars: Fusing deep learning and
  geometric modeling,'' in \emph{IEEE Intelligent Vehicles Symposium
  (IV)}.\hskip 1em plus 0.5em minus 0.4em\relax IEEE, 2017, pp. 1025--1032.

\bibitem{li2017traffic}
L.~Li, B.~Qian, J.~Lian, W.~Zheng, and Y.~Zhou, ``Traffic scene segmentation
  based on rgb-d image and deep learning,'' \emph{IEEE Transactions on
  Intelligent Transportation Systems}, vol.~19, no.~5, pp. 1664--1669, 2017.

\bibitem{hadsell2008deep}
R.~Hadsell, A.~Erkan, P.~Sermanet, M.~Scoffier, U.~Muller, and Y.~LeCun, ``Deep
  belief net learning in a long-range vision system for autonomous off-road
  driving,'' in \emph{IEEE/RSJ International Conference on Intelligent Robots
  and Systems (IROS)}.\hskip 1em plus 0.5em minus 0.4em\relax IEEE, 2008, pp.
  628--633.

\bibitem{iso_2022}
\textit{Road Vehicles — Safety of the Intended Functionality}, ISO/PAS
  21448:2019, International Organization for Standardization, Geneva,
  Switzerland, June 2022.

\bibitem{guo2020recognizing}
Z.~Guo, Y.~Pan, G.~Zhao, J.~Zhang, and N.~Dong, ``Recognizing hazard perception
  in a visual blind area based on eeg features,'' \emph{IEEE Access}, vol.~8,
  pp. 48\,917--48\,928, 2020.

\bibitem{peng2022application}
Y.~Peng, Q.~Xu, S.~Lin, X.~Wang, G.~Xiang, S.~Huang, H.~Zhang, and C.~Fan,
  ``The application of electroencephalogram in driving safety: current status
  and future prospects,'' \emph{Frontiers in Psychology}, vol. 13: 919695,
  2022.

\bibitem{li_chang_sui_2022}
H.~Li, R.~Chang, and X.~Sui, ``The effect of the degree and location of danger
  in traffic hazard perception: An erp study,'' \emph{NeuroReport}, vol.~33,
  no.~5, pp. 215--220, 2022.

\bibitem{li2022exploratory}
X.~Li, L.~Yang, and X.~Yan, ``An exploratory study of drivers’ eeg response
  during emergent collision avoidance,'' \emph{Journal of Safety Research},
  vol.~82, pp. 241--250, 2022.

\bibitem{ma_bai_pei_xu_2018}
Q.~Ma, X.~Bai, G.~Pei, and Z.~Xu, ``The hazard perception for the surrounding
  shape of warning signs: Evidence from an event-related potentials study,''
  \emph{Frontiers in Neuroscience}, vol. 12: 824, 2018.

\bibitem{zhu_ma_bai_hu_2020}
L.~Zhu, Q.~Ma, X.~Bai, and L.~Hu, ``Mechanisms behind hazard perception of
  warning signs: An eeg study,'' \emph{Transportation Research Part F: Traffic
  Psychology and Behaviour}, vol.~69, pp. 362--374, 2020.

\bibitem{guy2016cortical}
M.~W. Guy, N.~Zieber, and J.~E. Richards, ``The cortical development of
  specialized face processing in infancy,'' \emph{Child Development}, vol.~87,
  no.~5, pp. 1581--1600, 2016.

\bibitem{polezzi2008predicting}
D.~Polezzi, L.~Lotto, I.~Daum, G.~Sartori, and R.~Rumiati, ``Predicting
  outcomes of decisions in the brain,'' \emph{Behavioural Brain Research}, vol.
  187, no.~1, pp. 116--122, 2008.

\bibitem{de1999effect}
V.~De~Pascalis, J.~Strelau, and B.~Zawadzki, ``The effect of temperamental
  traits on event-related potentials, heart rate and reaction time,''
  \emph{Personality and Individual Differences}, vol.~26, no.~3, pp. 441--465,
  1999.

\bibitem{gan2021constructing}
S.~Gan, Q.~Li, Q.~Wang, W.~Chen, D.~Qin, and B.~Nie, ``Constructing
  personalized situation awareness dataset for hazard perception,
  comprehension, projection, and action of drivers,'' in \emph{IEEE
  International Intelligent Transportation Systems Conference (ITSC)}.\hskip
  1em plus 0.5em minus 0.4em\relax IEEE, 2021, pp. 1697--1704.

\bibitem{dosovitskiy2017carla}
A.~Dosovitskiy, G.~Ros, F.~Codevilla, A.~Lopez, and V.~Koltun, ``{CARLA: An
  open urban driving simulator},'' in \emph{Conference on Robot Learning
  (CoRL)}.\hskip 1em plus 0.5em minus 0.4em\relax PMLR, 2017, pp. 1--16.

\bibitem{eprime}
\BIBentryALTinterwordspacing
\textit{E-Prime 3.0.} {(2016).} {Psychology Software Tools, Inc.} Accessed:
  Aug. 1, 2022. [Online]. Available: \url{https://support.pstnet.com/}
\BIBentrySTDinterwordspacing

\bibitem{delorme2004eeglab}
A.~Delorme and S.~Makeig, ``Eeglab: An open source toolbox for analysis of
  single-trial eeg dynamics including independent component analysis,''
  \emph{Journal of Neuroscience Methods}, vol. 134, no.~1, pp. 9--21, 2004.

\bibitem{kataoka2018drive}
H.~Kataoka, T.~Suzuki, S.~Oikawa, Y.~Matsui, and Y.~Satoh, ``Drive video
  analysis for the detection of traffic near-miss incidents,'' in \emph{IEEE
  International Conference on Robotics and Automation (ICRA)}.\hskip 1em plus
  0.5em minus 0.4em\relax IEEE, 2018, pp. 3421--3428.

\bibitem{kataoka2020joint}
H.~Kataoka, T.~Suzuki, K.~Nakashima, Y.~Satoh, and Y.~Aoki, ``Joint pedestrian
  detection and risk-level prediction with
  motion-representation-by-detection,'' in \emph{IEEE International Conference
  on Robotics and Automation (ICRA)}.\hskip 1em plus 0.5em minus 0.4em\relax
  IEEE, 2020, pp. 1021--1027.

\bibitem{geiger2012kitti}
A.~Geiger, P.~Lenz, and R.~Urtasun, ``Are we ready for autonomous driving? the
  kitti vision benchmark suite,'' in \emph{IEEE Conference on Computer Vision
  and Pattern Recognition (CVPR)}.\hskip 1em plus 0.5em minus 0.4em\relax IEEE,
  2012, pp. 3354--3361.

\bibitem{cordts2016cityscapes}
M.~Cordts, M.~Omran, S.~Ramos, T.~Rehfeld, M.~Enzweiler, R.~Benenson,
  U.~Franke, S.~Roth, and B.~Schiele, ``The cityscapes dataset for semantic
  urban scene understanding,'' in \emph{IEEE Conference on Computer Vision and
  Pattern Recognition (CVPR)}.\hskip 1em plus 0.5em minus 0.4em\relax IEEE,
  2016, pp. 3213--3223.

\bibitem{koechlin2007anterior}
E.~Koechlin and A.~Hyafil, ``Anterior prefrontal function and the limits of
  human decision-making,'' \emph{Science}, vol. 318, no. 5850, pp. 594--598,
  2007.

\bibitem{gilbert2006functional}
S.~J. Gilbert, S.~Spengler, J.~S. Simons, J.~D. Steele, S.~M. Lawrie, C.~D.
  Frith, and P.~W. Burgess, ``Functional specialization within rostral
  prefrontal cortex (area 10): A meta-analysis,'' \emph{Journal of Cognitive
  Neuroscience}, vol.~18, no.~6, pp. 932--948, 2006.

\bibitem{fincham2002neural}
J.~M. Fincham, C.~S. Carter, V.~Van~Veen, V.~A. Stenger, and J.~R. Anderson,
  ``Neural mechanisms of planning: A computational analysis using event-related
  fmri,'' \emph{Proceedings of the National Academy of Sciences}, vol.~99,
  no.~5, pp. 3346--3351, 2002.

\bibitem{goel1995modeling}
V.~Goel, J.~Grafman, N.~Sadato, M.~Hallett, \emph{et~al.}, ``Modeling other
  minds,'' \emph{NeuroReport}, vol.~6, no.~13, pp. 1741--1746, 1995.

\bibitem{goel1997seats}
V.~Goel, B.~Gold, S.~Kapur, and S.~Houle, ``The seats of reason? {A}n imaging
  study of deductive and inductive reasoning,'' \emph{NeuroReport}, vol.~8,
  no.~5, pp. 1305--1310, 1997.

\bibitem{volz2005variants}
K.~G. Volz, R.~I. Schubotz, and D.~Y. von Cramon, ``Variants of uncertainty in
  decision-making and their neural correlates,'' \emph{Brain Research
  Bulletin}, vol.~67, no.~5, pp. 403--412, 2005.

\bibitem{gharib2020neural}
S.~Gharib, A.~Zare-Sadeghi, S.~A. Zakerian, and M.~R. Haidari, ``The neural
  basis of hazard perception differences between novice and experienced
  drivers-an fmri study,'' \emph{Excli Journal}, vol.~19, pp. 547--566, 2020.

\end{thebibliography}

\end{document}